# AN ALGORITHM FOR ODD GRACEFUL LABELING OF THE UNION OF PATHS AND CYCLES

M. Ibrahim Moussa

Faculty of Computers & Information, Benha University, Benha, Egypt
moussa_6060@yahoo.com

## ABSTRACT

*In 1991, Gnanajothi [4] proved that the path graph $P_n$ with $n$ vertex and $n-1$ edge is odd graceful, and the cycle graph $C_m$ with $m$ vertex and $m$ edges is odd graceful if and only if $m$ even, she proved the cycle graph is not graceful if $m$ odd. In this paper, firstly, we studied the graph $C_m \cup P_n$ when $m = 4, 6, 8, 10$ and then we proved that the graph $C_m \cup P_n$ is odd graceful if $m$ is even. Finally, we described an algorithm to label the vertices and the edges of the vertex set $V(C_m \cup P_n)$ and the edge set $E(C_m \cup P_n)$.*

## 1. INTRODUCTION

The study of graceful graphs and graceful labeling methods was introduced by Rosa [1]. Rosa defined a β- valuation of a graph $G$ with $q$ edges an injection from the vertices of $G$ to the set $\{0,1,2…q\}$ such that when each edge $uv$ is assigned the label $|f(u) - f(v)|$, the resulting edges are distinct. β- Valuation is a function that produces graceful labeling. However, the term graceful labeling was not used until Golomb studied such labeling several years later [2]. A graph $G$ of size $q$ is odd-graceful, if there is an injection $f$ from $V(G)$ to $\{0,1,2…2q-1\}$ such that, when each edge $uv$ is assigned the label or weight $|f(u) - f(v)|$, the resulting edge labels are $\{1,3,5…2q-1\}$. This definition was introduced by Gnanajothi [4] in 1991. In 1991, Gnanajothi [4] proved the graph $C_m \times K_2$ is odd-graceful if and only if $m$ even. She also proved that the graph obtained from $P_n \times P_2$ by deleting an edge that joins to end points of the $P_n$ paths and this last graph knew as the ladder graph. She proved that every graph with an odd cycle is not odd graceful. She also proved the following graphs are odd graceful: $P_n$; $C_m$ if and only if $m$ is even and the disjoint union of copies of $C_4$. In 2000, Kathiresan [6] used the notation $P_{n;m}$ to denote the graph obtained by identifying the end points of $m$ paths each one has length $n$. In 1997 Eldergill [5] generalized Gnanajothi [4] result on stars by showing that the graph obtained by joining one end point from each of any odd number of paths of equal length is odd graceful graph. In 2002 Sekar [7] proved that the graphs; $P_{n;m}$ when $n \geq 2$ and $m$ is odd, $P_{2;m}$ and $m \geq 2$, $P_{4;m}$ and $m \geq 2$, $P_{n;m}$ when $n$ and $m$ are even and $n \geq 4$ and $m \geq 4$, $P_{4r+1;4r+2}$, $P_{4r-1;4r}$, and all $n$-polygonal snakes with $n$ even are odd graceful. In 2009 Moussa [9] presented some algorithms to prove for all $m \geq 2$ the following graphs $P_{4r-1;m}$, $r =1,2,3$ and $P_{4r+1;m}$, $r =1,2$ are odd graceful. He presented algorithm which show that the closed spider graph and the graphs obtained by joining one or two paths $P_m$ to each vertex of the path $P_n$ are odd graceful. He used the cycle representation and denoted £ -representation to present a simple labeling graph algorithm, the cycle representation is similar to the $\pi$ -representation made by Kotazig [3]. In 2009 Moussa and Bader [8] have presented the algorithms that showed the graphs obtained by joining $n$ pendant edges to each vertex of $C_m$ are odd graceful if and only if $m$ is even.

In this paper we show that $C_m \cup P_n$ is odd graceful if $m$ is even and $n > m-2$, if the number of edges in the cycle $C_m$ can be equally divisible by four, and $n > m-4$ for all other even value of $m$. We first explicitly define an odd graceful labeling of $C_4 \cup P_n$, $C_6 \cup P_n$, $C_8 \cup P_n$, and $C_{10} \cup P_n$ and then, using this odd graceful labeling, describe a recursive procedure to obtain an odd graceful labeling of $C_m \cup P_n$. Finally; we present an algorithm for computing the odd graceful labeling of the union of path and cycle graphs, we prove the correctness of the algorithms at the end of this paper. The remainder of this paper is organized as follows. In section 2, illustrate the need of graph labeling and we mention the existing variety types of labeling methods. In section 3 we give some assumptions and definitions related with the odd graceful labeling and graphs. In section 4, we present and discuss the odd label of union of paths and cycle. Section 5 is the conclusion of this research.

## 2. RELATED WORK.

The graph labeling serves as useful models for a broad range of applications such as: radar, communications network, circuit design, coding theory, astronomy, x-ray, crystallography, data base management and models for constraint programming over finite domains. J. Gallian in his dynamic survey [11], he has collected everything on graph labeling, he observed that over thousand papers have been studied and many kinds of graph labeling have been defined, viz.: Graceful Labeling, Harmonious Labeling, Magic Labelings, balanced labeling, $k$ -graceful Labeling, γ-labeling, and Odd-Graceful Labelings. For further information about the graph labeling, we advise the reader to refer to the brilliant dynamic survey on the subject [11].

## 3. ASSUMPTIONS AND DEFINITIONS.

**Definition 1[4]**
Let $G$ be a finite simple graph, whose vertex set is denoted $V(G)$, while $E(G)$ denotes its edge set, the order of $G$ is the cardinality $n = |V(G)|$ and the size of $G$ is the cardinality $q = |E(G)|$. We write $uv \in E(G)$ if there is an edge connecting the vertices $u$ and $v$ in $G$. An odd graceful labeling of a graph $G$ is a one to one function $f : V(G) \to \{0,1,2,...,2q-1\}$. Such that, when each edge $uv$ is assigned the label $f^*(uv) = |f(u) - f(v)|$ the resulting edge labels are $\{0, 1, 2 ... 2q - 1\}$.

**Definition 2[10]**

A path in a graph is a sequence of vertices such that from each of its vertices there is an edge to the next vertex in the sequence. The first vertex is called the start vertex and the last vertex is called the end vertex. Both of them are called end or terminal vertices of the path. The other vertices in the path are internal vertices. A cycle is a graph with an equal number of vertices and edges whose vertices can be placed around a circle so that two vertices are adjacent if and only if they appear consecutively along the circle. The graph has $n$ or $m$ vertex that is a path or a cycle is denoted $P_n$ or $C_m$, respectively. The union of two graphs $G_1 = (V_1, E_1)$ and $G_2 = (V_2, E_2)$, written $G_1 \cup G_2$, is the graph with vertex set $V(G_1 \cup G_2) = V(G_1) \cup V(G_2)$ and the edge set $E(G_1 \cup G_2) = E(G_1) \cup E(G_2)$.

## 4. VARIATION OF ODD GRACEFUL LABELING

### 4.1. Odd gracefulness of $C_4 \cup P_n$

**Theorem1** $C_4 \cup P_n$ is odd graceful for every integer $n > 2$.

**Proof**

Let $V(C_4) = \{u_1, u_2, u_3, u_4\}$, $V(P_n) = \{v_1, ..., v_n\}$ where $V(C_4)$ is the vertex set of the cycle $C_4$ and $V(P_n)$ is the vertex set of the path $P_n$, and $q = n + 3$, see Fig.1. For every vertex $u_i$ and $v_i$, the odd graceful labeling functions $f(u_i)$ and $f(v_i)$ respectively as follows

$$f(u_1) = 0, \; f(u_2) = 2q - 1, \; f(u_3) = 2, \; f(u_4) = 2q - 5 \text{ and}$$

$$f(v_i) = \begin{cases} i & i \text{ odd} \\ 2q - i - 6 & i \text{ even} \end{cases}$$

The edge labeling function $f^*$ defined as follows:

$f^*(u_1 u_2) = 2q - 1$, $f^*(u_2 u_3) = 2q - 3$, $f^*(u_3 u_4) = 2q - 7$, $f^*(u_4 u_1) = 2q - 5$, and

$f^*(v_i v_{i+1}) = 2q - 2i - 7 \quad i = 1, 2, ..., n - 1$

Figure1 shows the method labeling of the graph $C_4 \cup P_n$ this complete the proof.∎

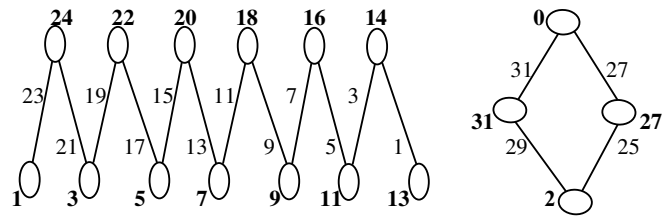

Figure 1.

### 4.2. Odd gracefulness of $C_6 \cup P_n$

**Theorem2** $C_6 \cup P_n$ is odd graceful for every integer $n > 2$.

**Proof**

Let $V(C_6) = \{u_1, u_2, u_3, u_4, u_5, u_6\}$, $V(P_n) = \{v_1, ..., v_n\}$ where $V(C_6)$ is the vertex set of the cycle $C_6$ and $V(P_n)$ is the vertex set of the path $P_n$, and $q = n + 5$, see Fig. 2. For every vertex $u_i$ and $v_i$, the odd graceful labeling functions $f(u_i)$ and $f(v_i)$ respectively as follows:

$f(u_1) = 0, f(u_2) = 2q-1, f(u_3) = 2, f(u_4) = 2q-3, f(u_5) = 4, f(u_6) = 2q-9$

and $f(v_i) = \begin{cases} i+2 & 3 \leq i \text{ odd} \\ i & i=1 \\ 2q-12+i & i \text{ even} \end{cases}$

The edge labeling function $f^*$ defined as follows:

$f^*(u_1u_2) = 2q-1, f^*(u_2u_3) = 2q-3, f^*(u_3u_4) = 2q-5, f^*(u_4u_5) = 2q-7,$

$f^*(u_5u_6) = 2q-13$, and $f^*(u_6u_1) = 2q-9$

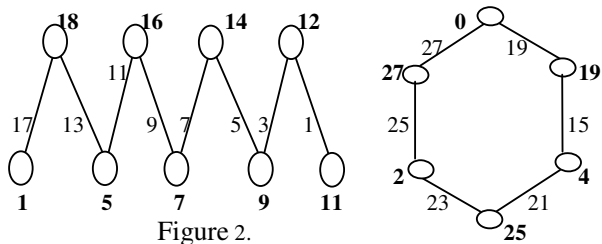

Figure 2.

## 4.3. Odd gracefulness of $C_8 \cup P_n$

**Theorem 3** $C_8 \cup P_n$ is odd graceful if and only if $n \geq 7$.

**Proof**:

For every vertex $u_i$ and $v_i$ in $V(P_n \cup C_8)$, we defined the odd graceful labeling functions $f(u_i)$ and $f(v_i)$ respectively as follows

$f(u_1) = 0, f(u_2) = 2q-1, f(u_3) = 2, f(u_4) = 2q-3, f(u_5) = 4, f(u_6) = 2q-5, f(u_7) = 6,$

$f(u_8) = 2q-13$, and $f(v_i) = \begin{cases} i & i \text{ odd} \\ 2q-14 & i=2 \\ 2q-i-14 & 4 \leq i \text{ even} \end{cases}$

And the function $f^*$ is defined as follows:

$f^*(u_1u_2) = 2q-1, f^*(u_2u_3) = 2q-3, f^*(u_3u_4) = 2q-5, f^*(u_4u_5) = 2q-7,$

$f^*(u_5u_6) = 2q-9, f^*(u_6u_7) = 2q-11, f^*(u_7u_8) = 2q-19, f^*(u_8u_1) = 2q-13$

$f^*(v_1v_2) = 2q-15, f^*(v_2v_3) = 2q-17, f^*(v_iv_{i+1}) = 2q-2i-15, \quad i = 3, ..., q-8$

The labeling of the graph $C_8 \cup P_{12}$ is indicated by Fig.3. ∎

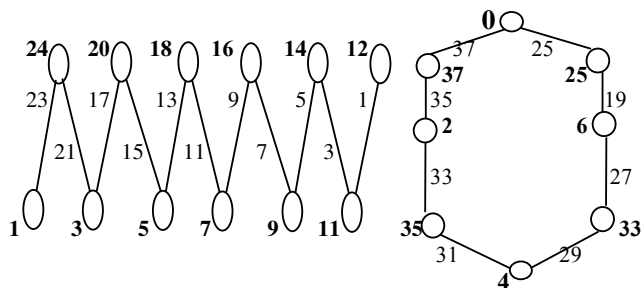

Figure 3.

### 4.4. Odd gracefulness of $C_{10} \cup P_n$

**Theorem 4** $C_{10} \cup P_n$ is odd graceful for every integer $n \geq 6$.

**Proof:**

For every vertex $u_i$ and $v_i$, we defined the odd graceful labeling functions $f(u_i)$ and $f(v_i)$ respectively as follows:

$f(u_1) = 0$, $f(u_2) = 2q - 1$, $f(u_3) = 2$, $f(u_4) = 2q - 3$, $f(u_5) = 4$, $f(u_6) = 2q - 5$,

$f(u_7) = 6$, $f(u_8) = 2q - 7$, $f(u_9) = 8$, $f(u_{10}) = 2q - 17$, and

$$f(v_i) = \begin{cases} i + 2 & 5 \leq i \text{ odd} \\ i & i = 1, 3 \\ 2q - i - 16 & i \text{ even} \end{cases}$$

So we obtain all the edge labels and the function $f^*$ is defined as follows:

$f^*(u_1 u_2) = 2q - 1$, $f^*(u_2 u_3) = 2q - 3$, $f^*(u_3 u_4) = 2q - 5$, $f^*(u_4 u_5) = 2q - 7$

$f^*(u_5 u_6) = 2q - 9$, $f^*(u_6 u_7) = 2q - 11$, $f^*(u_7 u_8) = 2q - 13$,

$f^*(u_8 u_9) = 2q - 15$, $f^*(u_9 u_{10}) = 2q - 25$, $f^*(u_{10} u_1) = 2q - 17$, and

$f^*(v_1 v_2) = 2q - 19$, $f^*(v_2 v_3) = 2q - 21$, $f^*(v_3 v_4) = 2q - 23$,

$f^*(v_i v_{i+1}) = 2q - 2i - 19$, $i = 4, ..., q - 10$.

### 4.5. Odd gracefulness of $C_m \cup P_n$

**Theorem 5**
Let $k$ is a given integer and $m = 2k$, the graph $C_m \cup P_n$ is odd graceful for every $n > m - 2$, $k$ is even, if $k$ is odd number the graph $C_m \cup P_n$ is odd graceful for every $n > m - 4$.

**Proof:**

Let $V(C_m) = \{u_1, ..., u_m\}$, $V(P_n) = \{v_1, ..., v_n\}$, where $V(C_m)$ is the vertex set of the cycle $C_m$ and $V(P_n)$ is the vertex set of the path $P_n$, and $q = n + m - 1$. For every vertex $u_i$ and $v_i$, we defined the odd graceful labeling functions $f(u_i)$ and $f(v_i)$ respectively as follows:

$f(u_1) = 0$, $f(u_2) = 2q - 1$, $f(u_3) = 2$, $f(u_4) = 2q - 3$,

$f(u_5) = 4$, $f(u_6) = 2q - 5$, $f(u_9) = 8$, ..., $f(u_m) = 2q - (2m - 3)$

If the value $m = 2k$, $k$ is odd number, the vertex $v_2$ would be labeled $f(v_2) = 2q - 2m + 2$ which decreased by two at every new value $i = 4, 6, ..., k - 2$, this means that

$f(v_i) = f(v_{i-2}) - 2 = 2q - 2m - (i - 4)$, and

$$f(v_i) = \begin{cases} i + 2 & k \leq i \text{ odd} \\ i & i = 1, 3, ..., k - 2 \\ 2q - 2m - (i - 4) & i \text{ even} \end{cases}$$

If $m=2k$, $k$ is even number, the vertex $v_2$ would be labeled $f(v_2) = 2q - 2m + 2$ which decreased by two at every new value $i = 4, 6, ..., k-2$. For $i = k-2$ the label value is $f(v_{k-2}) = 2q - 2m + 6 - k$ while the label value of the vertex $v_k$ is four out of the value $f(v_{k-2})$, this means that

$f(v_{i=k}) = f(v_{k-2}) - 4 = 2q - 2m + 2 - i$, and

$$f(v_i) = \begin{cases} i & i \; odd \\ 2q - 2m + 4 - i & i = 2, 4, ..., k-2 \\ 2q - 2m + 2 - i & k \leq i \; even \end{cases}$$

The function $f^*$ induces the edge labels of the cycle $C_m$ as the following:

$f^*(u_1 u_2) = 2q - 1, \; f^*(u_2 u_3) = 2q - 3, \; f^*(u_3 u_4) = 2q - 5, ...,$

$f^*(u_{m-1} u_m) = 2q - 3m + 5, \; f^*(u_m u_1) = 2q - 2m + 3$.

Function $f^*$ induces the edge labels of the path as follows:

$f^*(v_1 v_2) = 2q - 2m + 1, \; f^*(v_2 v_3) = 2q - 2m - 1, ..., f^*(v_{k-2} v_{k-1}) = 2q - 3m + 7,$

$f^*(v_{k-1} v_k) = 2q - 3m + 3, ......., f^*(v_k v_{k+1}) = 2q - 3m + 1, ........., 1$

There is a guarantee that each component in the given graph has odd graceful, the path graph is odd graceful, the cycle graph with an even number of vertices is odd graceful (see [4]). We have to prove that the vertex labels are distinct and all the edge labels are distinct odd numbers $\{1, 3, 5...2q-1\}$. The edge labels of $C_m$ are numbered according to the decreasing sequence $2q-1, 2q-3, .......$ . The edge labels of $P_n$ are numbered according to the decreasing sequence $f^*(v_i v_{i+1}) = 2q - 2i - (2m-1), \; i = 4, ..., q-m$. The reader can easily find out, if $i = q-m$ the last edge label is equal one; this means that the edge labels take values in $\{2q-1, 2q-3, ..., 1\}$. In order to prevent any vertex in $P_n$ to share label with a vertex in $C_m$, the difference between the largest even label and the smallest even label in $P_n$ have to be more than the largest even label in $C_m$, this leads to two cases:

Case I: if $m = 2k$, $k$ is even then $(f(u_m) - 1) - 2\lceil n/2 \rceil > m - 2 \Rightarrow n > m - 2$

Case I: if $m = 2k$, $k$ is odd then $(f(u_m) - 1) - 2(\lceil n/2 \rceil - 1) > m - 2 \Rightarrow n > m - 4$. ∎

### 4.6. The proposed sequential algorithm

The union graph $P_n \cup C_m$ has a vertex set $V(P_n \cup C_m) = V(P_n) \cup V(C_m)$ with cardinality $n+m$ and an edge set $E(P_n \cup C_m) = E(P_n) \cup E(C_m)$ with cardinality $q = m + n - 1$. Let the cycle $C_m$ is demonstrated by listing the vertices and the edges in the order $u_1, e_1, u_2, e_2, ..., u_{m-1}, e_{m-1}, u_m, e_m, u_1$. We name the vertex $u_m$ ACTIVE vertex, the vertex $u_m$ is an endpoint of the edge $e_{m-1}$, and we name the edge $e_{m-1}$ DOUBLE-JUMP edge. The path $P_n$ is demonstrated by listing the vertices and the edges in the order $v_1, e'_1, v_2, e'_2, ..., v_{n-1}, e'_{n-1}, v_n$. The algorithm has two passes; they can run in a sequential or a parallel way and it works in a way similar to the above labeling in section 3.5. In one pass, the algorithm labels the vertices and the edges in the cycle $C_m$. For the other pass, it labels the vertices and the edges of the path $P_n$. At the beginning of the algorithm, we are computing the odd label function for the ACTIVE vertex and the DOUBLE-JUMP edge. The ACTIVE vertex has the odd graceful labeling function $f(u_m) = 2q - (2m-3)$, the vertex $u_m$ has the smallest odd label value between the vertices in the cycle $C_m$. The DOUBLE-JUMP edge is assigned the label function $f^*(e_{m-1}) = 2q - 3m + 5$. The given label to the ACTIVE vertex and the DOUBLE-JUMP edge computed independently from other vertices or edges in the graph.

1. Number the ACTIVE vertex with the value $f(u_m) = 2q - (2m-3)$
2. Number the DOUBLE-JUMP edge with the value $f^*(e_{m-1}) = 2q - 3m + 5$

Algorithm 1: Procedure Initialization

In the first pass, the algorithm starts at the vertex $u_1$, there are two main steps that can be performed. These steps (in particular order) are: performing an action on the current vertex (referred to as "numbering" the vertex), number the current vertex with the value $f(u_1) = 0$, traversing to the left adjacent vertex $u_2$ and number it with the value $f(u_2) = 2q - 1$, and traversing to the left adjacent vertex $u_3$ and number it with the value $f(u_3) = 2$ traversing to the left adjacent vertex $u_4$ and number it with the value $f(u_4) = 2q - 3$, traversing to the left adjacent vertex $u_5$ and number it with the value $f(u_5) = 4$ .. Thus the process is most easily described through recursion. Finally, reach to the ACTIVE vertex which has the exception label and number it with the value $f(u_m) = 2q - (2m-3)$, the edge's labeling induced by the absolute value of the difference of the vertex's labeling. To label the cycle $C_m$ odd graceful, perform the following operations, starting with $u_1$:

1. Number the vertex $u_1$ with the value $f(u_1) = 0$
2. For ( $i = 3$; $i \leq m-2$; $i += 2$ )
    $f(u_i) = f(u_{i-2}) + 2$
3. For ( $i = 2$; $i \leq m-1$; $i += 2$ )
    $f(u_i) = 2q - i + 1$
4. Number the ACTIVE vertex with $f(u_m) = 2q - (2m-3)$.
5. Compute the edge labels by taking the absolute value of the difference of incident vertex labels.

Algorithm 2: Odd graceful labeling of $C_m$

After the above process, the algorithm starts the second pass to label the vertices and edges of the path component $P_n$. Second pass starts at the edge $e'_1 = (v_1, v_2)$, its label value is $f^*(e'_1) = f(u_m) - 2$, if the label value of the edge $e'_1$ equals to the label value of the DOUBLE-JUMP edge renumber it with the value $f^*(e'_1) = f^*(e'_1) - 2$ and number the vertex $v_1$ with the label value $f(v_1) = 1$, traversing to the vertex $v_2$ and number it with the value $f(v_2) = f^*(e'_1) + 1$. Traversing to the next incident edge $e'_2$ and number it with the value $f^*(e'_2) = f^*(e'_1) - 2$ if the label value of the edge $e'_2$ equals to the label value of the DOUBLE-JUMP edge renumber it with the value $f^*(e'_2) = f^*(e'_2) - 2$, traverse to the next vertex $v_3$ which induces the label value $f(v_3) = f(v_2) - f^*(e'_2)$, otherwise traverse to the next vertex $v_3$, without double subtracting for the label value of the edge $e'_2$, and number it with the value $f(v_3) = f(v_2) - f^*(e'_2)$, traverse to the next vertex $v_4$ which induces the label value $f(v_4) = f(v_3) + f^*(e'_3)$. Thus the process is most easily described through recursion again. To label the path $P_n$ odd graceful labeling, perform the following operations, starting with the edge $e'_1 = (v_1, v_2)$:

1. *Number the vertex $v_1$ with the value $f(v_1) = 1$*
2. *Number an auxiliary edge $e'_0$ with $f^*(e'_0) = f(u_m)$*
3. *For ( $j = 1$; $j \leq n$-1; $j$ += 1 )*
   3.1 Number the edge $e'_j$ with $f^*(e'_j) = f^*(e'_{j-1}) - 2$
   3.2 If ( $f^*(e'_j) = f^*(e_{m-1})$ ) Renumber the edge $e'_j$ with the value $f^*(e'_j) = f^*(e'_j) - 2$
   3.3 Number the vertex $v_{j+1}$ with the value $f(v_{j+1}) = f^*(e'_j) + (-)^{j+1} f(v_j)$
4. *End For*

Algorithm 3: Odd graceful labeling of $P_n$

The algorithm is traversed exactly once for each vertex and edge in the graph $P_n \cup C_m$, since the size of the graph equals $q$ then at most O ($q$) time is spent in total labeling of the vertices and edges, thus the total running time of the algorithm is O ($q$). The parallel algorithm for the odd graceful labeling of the graph $P_n \cup C_m$, based on the above proposed sequential algorithm is building easily. Since all the above three subroutine are independent and there is no reason to sort their executing out, so they are to join up parallel in the same time point.

## 5. CONCLUSION

In this paper, we first explicitly defined an odd graceful labeling of $C_4 \cup P_n$, $C_6 \cup P_n$, $C_8 \cup P_n$, and $C_{10} \cup P_n$ and then using this odd graceful labeling to have generalized results by describing a recursive procedure to obtain an odd graceful labeling of $C_m \cup P_n$, if $m$ is even and $n > m - 2$, if the number of edges in the cycle $C_m$ can be equally divisible by four, and $n > m - 4$ for all other even value of $m$. After we introduced a general form for labeling the union of the paths and the cycles in odd graceful label, we described a sequential algorithm to label the vertices and the edges of the graph $P_n \cup C_m$. The sequential algorithm runs in linear with total running time equals O ($q$). The parallel version of the proposed algorithm, as we showed, existed and it is described shortly.